\documentclass{article}
\usepackage{amsmath}
\usepackage{psfrag}
\usepackage[dvips]{graphicx}
\newcommand{\D}{\,\mathrm{d}}

\newcommand{\pdl}[2]{\ensuremath{\partial #1 / \partial #2}}
\newcommand{\pdc}[3]{\ensuremath{\left(\frac{\partial #1}{\partial #2}\right)_{#3}}}
\newcommand{\pdcl}[3]{\ensuremath{\left(\partial #1 / \partial #2\right)_{#3}}}
\newcommand{\jac}[8]{\ \begin{array}{@{\:}r@{}c@{\ }c@{\ }c@{\ }c@{}r@{\:}}\partial(& #1, & #2, & #3, & #4 &)\\\hline\partial(& #5, & #6, & #7, & #8 &)\end{array}\ }
\newcommand{\ie}{\textit{i.e.,}~}
\newcommand{\eg}{\textit{e.g.,}~}
\newcommand{\vs}{\ensuremath{v_\text{s}}}
\newcommand{\vsb}{\ensuremath{\mathbf{v}_\text{s}}}
\newcommand{\vnb}{\ensuremath{\mathbf{v}_\text{n}}}
\renewcommand{\j}{\ensuremath{\mathbf{j}}}

\begin{document}
\psfragscanon

\author{A.F.\,Andreev\footnote{E-mail: andreev@kapitza.ras.ru},
L.A.\,Melnikovsky\footnote{E-mail: leva@kapitza.ras.ru}}

\title{Thermodynamic inequalities in superfluid}
\date{\today}
\maketitle
\vspace{-8mm}
\begin{center}
\textit{P.L.Kapitza Institute for Physical Problems}\\
\textit{Russian Academy of Sciences,}\\
\textit{Kosygin str., 2, 119334, Moscow, Russia}\\
\end{center}

\begin{abstract}
We investigate general thermodynamic stability conditions
for the superfluid. This analysis is performed in an extended
space of thermodynamic variables containing (along with the usual
thermodynamic coordinates such as pressure and temperature)
superfluid velocity and momentum density. The stability
conditions lead to {\em thermodynamic inequalities} which
replace the Landau superfluidity criterion
at finite temperatures.
\end{abstract}

%\noindent \textbf{PACS}{: 67.40.Bz, 67.40.Pm, 68.35.Ct, 61.72.Ji}

\section{Introduction}

Usually in experiments the vortices destroy superfluidity at velocities
far below the Landau critical velocity. This is why the superfluid
hydrodynamics equations can be expanded in powers of low velocities  and
one safely uses the first nontrivial terms of this expansion.

Nevertheless, there is a number of experiments (see \cite{exp}) where
the superfluid flow is investigated in small orifices. It has been shown
that in these circumstances the maximum velocity is a decreasing
function of the orifice width and may reach the order of the Landau
critical velocity if the aperture is small enough. This means that all
thermodynamic quantities of the superfluid become nontrivial functions
of the not small superfluid velocity (\ie it depends not only on the
usual thermodynamic coordinates such as pressure and temperature). The
only assumption one can make (and we do it) is that the fluid at rest is
isotropic. This quite general statement of the problem is used in the
paper; we find the complete set of thermodynamic inequalities in this
light, \ie the conditions imposed on thermodynamic functions for the
superfluid to remain stable. 

Finally we employ the Landau phonon-roton model to calculate the highest
velocity compatible with obtained thermodynamic inequalities and show
that it can be interpreted as a critical velocity. This thermodynamic
scenario supposedly explains the superfluidity break-up in small
orifices.

\section{Stability}
When deriving general superfluid hydrodynamic equations it
is usually supposed~\cite{khalat} that each infinitesimal
volume of the liquid is (locally) in equilibrium and this
equilibrium is stable. For the state of the liquid to be
stable, it should provide an entropy maximum (at least
local) for an isolated system. Instead of investigating the
condition of the entropy maximality, it is
convenient~\cite{LL5} to use another, equivalent to the
first one, condition, that is the condition of the energy
minimality under constant entropy and additive integrals of
motion. Thus, to examine if the state is stable or not, one
must investigate the second variation of the energy.
Such analysis will provide sufficient conditions for the
energy minimality.

Total energy of the superfluid $E_\text{tot}$ is an integral of the energy density $E$ over the
entire volume
\begin{equation}
E_\text{tot}=\int E \D \mathbf{r}.
\end{equation}
The energy density can be obtained via a Galilean
transformation
%is given by equation
\begin{equation}\label{Edensity}
E=\frac{\rho \vs^2}{2}+\vsb\j_0+E_0.
\end{equation}
Here \vsb\ is the superfluid velocity, $\rho$ is
the mass density and subscript $0$ denotes quantities
measured in the frame of reference of the superfluid
component (that is the frame where the superfluid velocity
is zero). Namely, $E_0$ and $\j_0$ are the energy
density and the momentum density (or, equally, the mass
flux) with respect to the superfluid component. The former is a
function of $\rho$, $\j_0$, and the entropy density
$S$. Its differential can be written as
\begin{equation}\label{E0density}
\D E_0= T \D S + \mu\D \rho +  \mathbf{w} \D \j_0,
\end{equation}
where Lagrange multipliers $T$, $\mu$, and $\mathbf{w}$ are
the temperature, the chemical potential, and the so-called
relative velocity of normal and superfluid components.
The liquid is isotropic and, consequently, the velocity $\mathbf{w}$
and the momentum density $\j_0$ are parallel to each other, as expressed by
\begin{equation*}
\j_0=j_0(T,\rho,w)\frac{\mathbf{w}}{w}.
\end{equation*}
This leads to a useful identity for the partial derivatives of $\j_0$ with
respect to $\mathbf{w}$:
\begin{equation} \label{djw}
\pdc{j_0^k}{w^l}{T,\rho}=\frac{w^k w^l}{w^2} \pdc{j_0}{w}{T,\rho}
+\left(\frac{\delta^{kl}}{w} - \frac{w^k w^l}{w^3}\right)j_0.
\end{equation}
Further transforming \eqref{Edensity}, we can rewrite it with the help of \eqref{E0density} in
the form
\begin{equation}
\D E= T\D S +
\left(\mu+\frac{\vs^2}{2}-\vsb\vnb\right)
\D\rho +(\j-\rho\vnb)\D\vsb +
\vnb\D \j,
\end{equation}
where we denoted the total momentum density
$\j=\rho\vsb + \j_0$ and the normal
velocity $\vnb=\vsb+\mathbf{w}$.

As usual, stability implies that each ``allowed'' fluctuation increases the total energy
of the system $E_\text{tot}$. Allowed are the fluctuations leaving conserved quantities
unchanged.
This means that the minimality of $E_\text{tot}$ must be investigated under
fixed entropy and all additive integrals of motion:
mass, momentum, and superfluid velocity.
While the conservation of mass and momentum is well-known, conservation of the
superfluid velocity
worths a special comment. Really, since the superfluid flow is irrotational, the velocity $\vsb$
is a gradient of a scalar: $\vsb=\nabla \phi$.
The same is true for the time derivative $\dot{\mathbf{v}}_\text{s}=\nabla \dot{\phi}$. This formula
expresses the conservation of all three components of the vector
\begin{equation}
{\mathbf{V}}_\text{s}=\int {\mathbf{v}}_\text{s} \D \mathbf{r}.
\end{equation}

Consider a macroscopic fluctuation of all the variables $\delta S$, $\delta \rho$,
$\delta \vsb$, and $\delta \j$. They are conserved and this ensures that
the first variation of the total energy for a uniform system is identically zero
\begin{equation}
\delta E_\text{tot}=\int \left(
                                           \pdc{E}{S}{\rho,\vsb,\j}\delta S
                                          +\pdc{E}{\rho}{S,\vsb,\j}\delta \rho
                                          +\pdc{E}{\vsb}{S,\rho,\j}\delta \vsb
                                          +\pdc{E}{\j}{S,\rho,\vsb}\delta \j
                                       \right)\D \mathbf{r} \equiv 0.
\end{equation}
The minimality criterion must be obtained as the condition of
the positive definiteness of the second differential quadratic form. The matrix of this quadratic
form is a Jacobian matrix $8\times 8$:
\begin{equation}
Q=\left\|\jac{T}{\vnb}{\ \mu+\vs^2/2-\vsb\vnb}{\ \j-\rho\vnb}
      {S}{\j}{\rho}{\vsb}\right\|.
\end{equation}
Common rule states that it is positive definite
if all principal minors $M_1, M_2,\dots M_8$ in the top-left corner are
positive. We recursively test these minors:
%%%%%%%%%%%%%%%%%%%%%%%%%%%%%%%%%%%
\begin{itemize}
\item The first positivity condition
\begin{multline*}
M_1=\jac{T}{\j}{\rho}{\vsb}{S}{\j}{\rho}{\vsb}=
\jac{T}{\j}{\rho}{\vsb}{T}{\vnb}{\rho}{\vsb}
\jac{T}{\vnb}{\rho}{\vsb}{S}{\j}{\rho}{\vsb}=\\
\pdc{j_0}{w}{T,\rho}
\left( 
  \pdc{S}{T}{\rho,w}\pdc{j_0}{w}{T,\rho} - \pdc{j_0}{T}{\rho,w}^2
\right)^{-1}
%=\left(\pdc{S}{T}{\rho,w} - \pdc{j_0}{w}{T,\rho}^{-1} \pdc{j_0}{T}{\rho,w}^2\right)^{-1}
>0
\end{multline*}
corresponds to the usual requirement of the heat capacity positivity. It is shown below
that $\pdcl{j_0}{w}{T,\rho}>0$, hence the last inequality eventually becomes
\begin{equation}\label{M1}
\pdc{S}{T}{\rho,w}\pdc{j_0}{w}{T,\rho} - \pdc{j_0}{T}{\rho,w}^2>0.
\end{equation}

\item 
Positivity of the next group of minors is easily verified with the following transformation
\begin{equation}\label{Q1}
Q'=\left\|\jac{T}{\vnb}{\rho}{\vsb}{S}{\j}{\rho}{\vsb}\right\|=
\left\|\jac{T}{\j}{\rho}{\vsb}{S}{\j}{\rho}{\vsb}\right\|
\left\|\jac{T}{\vnb}{\rho}{\vsb}{T}{\j}{\rho}{\vsb}\right\|.
\end{equation}
Whether the minors $M_2, M_3, M_4$ are positive is determined by
the second multiplier in \eqref{Q1}. Required condition is therefore equivalent
to the positive definiteness of the matrix 
\begin{equation*}
\left\|\pdc{\j}{\vnb}{T,\rho,\vsb}\right\|^{-1}=\\
\left\|\pdc{\j_0}{\mathbf{w}}{T,\rho}\right\|^{-1}=
\begin{Vmatrix}
\pdcl{j_0}{w}{T,\rho} &0      &0\\
0                     &j_0/w  &0\\
0                     &0      &j_0/w
\end{Vmatrix}^{-1}.
\end{equation*}
Here we used \eqref{djw} and chosen the direction of the $\mathbf{w}$ vector as the first coordinate.
This adds to our collection two more inequalities
\begin{equation}\label{M2a}
\j_0\mathbf{w}\geq 0,
\end{equation}
\begin{equation}\label{M2b}
\pdc{j_0}{w}{T,\rho}>0.
\end{equation}

\item The same transformation applied to the biggest minors gives:
\begin{equation*}
Q=
Q'
\left\|\jac{T}{\vnb}{\mu+\vs^2/2-\vsb\vnb}{\j-\rho\vnb}{T}{\vnb}{\rho}{\vsb}\right\|=
Q' Q''.
\end{equation*}
Again, the minors $M_5,M_6, M_7, M_8$ correspond to nontrivial principal minors of $Q''$.
We use the thermodynamic identity to relate the chemical potential~$\mu$
and the conventional pressure~$p$
\begin{equation*}
\D\mu=\frac{\D p}{\rho}-\frac{S}{\rho}\D T +
\vsb\D\mathbf{w}.
\end{equation*}
This gives
\begin{equation*}
\pdc{\left(\mu+\vs^2/2-\vsb\vnb\right)}{\rho}{T,\vnb,\vsb}=
\pdc{\mu}{\rho}{T,\mathbf{w}}=\frac{1}{\rho}\pdc{p}{\rho}{T,w}.
\end{equation*}
The following is an explicit representation of $Q''$ sub-matrix corresponding
to a four-dimensional space
$\rho,\vs^x,\vs^y,\vs^z$; as before we let the $x$-axis run along $\mathbf{w}$ direction.
Using \eqref{djw} we obtain
\begin{equation*}
%Q''=
%\left\|\jac{T}{\vnb}{\mu+\vs^2/2-\vsb\vnb}{\j-\rho\vnb}{T}{\vnb}{\rho}{\vsb}\right\|=
\begin{Vmatrix}
\pdcl{p}{\rho}{T,w}/\rho &\pdcl{j_0}{\rho}{T,w}-w		&0 &0\\
\pdcl{j_0}{\rho}{T,w}-w  &\rho-\pdcl{j_0}{w}{T,\rho}	&0 &0\\
0 &0 &\rho-j_0/w &0\\
0 &0 &0          &\rho-j_0/w
\end{Vmatrix}
.
\end{equation*}

Appropriate inequalities are:
\begin{equation}\label{M3}
\pdc{p}{\rho}{T,w}>0,
\end{equation}
which is literally a generalized (to a non-zero inter-component velocity $w$)
positive compressibility requirement,
\begin{equation}\label{M4a}
j_0<w\rho,
\end{equation}
and
\begin{equation}\label{M4b}
\pdc{p}{\rho}{T,w}
\left(\rho-\pdc{j_0}{w}{T,\rho}\right)
-\rho\left(\pdc{j_0}{\rho}{T,w}-w\right)^2>0.
\end{equation}
\end{itemize}

Inequalities \eqref{M1}, \eqref{M2a}, \eqref{M2b},
\eqref{M3}, \eqref{M4a}, and \eqref{M4b} are 
sufficient conditions for the thermodynamic stability.

\section{Discussion}
In a ``stopped-normal-component'' arrangement, the mass flux $\mathbf{f}$ with respect to the normal
component may become more convenient than $\j_0$---the mass flux relative to the superfluid one.
The obvious relation between them $f=\rho w - j_0$ leads to the following reformulation of the inequalities:
\begin{equation}\label{F1}
\mathbf{fw}<0, \quad f<w\rho,
\end{equation}
\begin{equation}\label{F2}
0<\pdc{f}{w}{\rho,T}<\rho
\end{equation}
\begin{equation}\label{F3}
\pdc{S}{T}{\rho,w}\left(\rho-\pdc{f}{w}{T,\rho}\right) > \pdc{f}{T}{\rho,w}^2,
\end{equation}
\begin{equation}\label{F4}
\pdc{p}{\rho}{T,w}\pdc{f}{w}{\rho,T}>\rho\pdc{f}{\rho}{w,T}^2
\end{equation}

As a simple application of the derived inequalities, consider them at $w=0$.
From \eqref{F1}, \eqref{F2},  \eqref{F3}, and  \eqref{F4} we get
\begin{equation}
\pdc{S}{T}{\rho,w}>0, \quad 
\pdc{p}{\rho}{T,w}>0,
\end{equation}
\begin{equation}
\rho>\pdc{j_0}{w}{T,\rho}>0.
\end{equation}
Using conventional notation, last inequality reads in the limit $w\rightarrow 0$
\begin{equation}\label{lambdapoint}
\rho_\text{s}>0, \quad \rho_\text{n}>0.
\end{equation}

\section{Phonon-Roton model}
Here we provide a usage example of the stability criteria for real superfluid ${}^4\mathrm{He}$.
To calculate derivatives involved in the inequalities one take refuge
in the microscopic approach. Simple and clear Landau phonon-roton model works pretty well
in wide temperature and velocity ranges. We use this model
to calculate the contribution of these quasiparticles to the ``modified'' free
energy in the frame of reference of the superfluid component:
\begin{equation}
\tilde{\cal F}_{0}= {\cal F}_{0}-\mathbf{w j}_0.
\end{equation}
Differential of this potential is given by
\begin{equation}
\D\tilde{\cal F}_{0}=-S\D T - \j_0\D \mathbf{w}.
\end{equation}
The modified free energy is obtained from the excitation spectrum with a conventional formula
\begin{equation}
\tilde{\cal F}_{0}=
T\int\ln\left(1-\exp\left(\frac{\mathbf{pw}-\epsilon(p)}{T}\right)\right)\frac{\D\mathbf{p}}{(2\pi\hbar)^3}.
\end{equation}
We denoted the excitation energy $\epsilon(p)$, which is given for two branches by the expressions
\begin{equation}
\epsilon_\text{ph}(p)=cp,
\quad
\epsilon_\text{r}(p)=\Delta+\frac{(p-p_0)^2}{2m}.
\end{equation}
Here and below, subscripts distinguish the quantities related to phonons and rotons, $c$
is the sound velocity, $\Delta$ is the roton energy gap, $m$ is the effective mass, and
$p_0$ is the momentum at the roton minimum\footnote{Data taken from \cite{don76,don81}:
$\rho=0.145\,\text{g}/\text{cm}^3$,
$\Delta=8.7\,\text{K}$,
$m=0.16 m_\text{He}$,
$p_0=3.673\,10^{8}\,\text{g}^{-1/3}\rho^{1/3}\hbar$,
$c=23800\,\text{cm}/\text{s}$,
$\pdl{\Delta}{\rho}= -0.47\,10^{-14}\,\text{cm}^5\text{s}^{-2}$,
$\pdl{m}{\rho}= -0.45\,10^{-23}\,\text{cm}^3$,
$\pdl{c}{\rho}= 467\,10^{3}\,\text{cm}^4 \text{s}^{-1}\text{g}^{-1}$.
}.
A small dimensionless parameter $m\Delta / p_0^2 \sim 0.03 \ll 1$
ensures, \eg that the Landau critical velocity is determined by
$v_\text{L}=\Delta/p_0$.

When integrated, these dispersion laws give the following contributions to the free energy:
\begin{gather}
\tilde{\cal F}_{0,\text{ph}}=
-\frac{T^4\pi^2}{90\hbar^3c^3}\left(1-\frac{w^2}{c^2}\right)^{-2},\\
\tilde{\cal F}_{0,\text{r}}=-\frac{T^{5/2}m^{1/2}}{2^{1/2}\pi^{3/2}\hbar^3}
 %\left( m\cosh\frac{wp_0}{T}+
 \frac{p_0}{w}\sinh\frac{wp_0}{T}
 %\right)
 \exp\left(
 %\frac{mw^2}{2T}
 -\frac{\Delta}{T}\right)
 .
\end{gather}
One can obtain all\footnote{We neglect the quasiparticle contribution to
the pressure derivative because it is just a small correction to the
speed of sound.} thermodynamic variables by differentiating this
potential. Namely
\begin{equation*}
S=-\pdc{\tilde{\cal F}_0}{T}{w,\rho},
\end{equation*}
\begin{equation*}
j_0=-\pdc{\tilde{\cal F}_0}{w}{T,\rho}.
\end{equation*}

%\begin{equation*}
%\tilde{p}=\pdc{\tilde{\cal F}_0}{\rho}{T,w},
%\end{equation*}
%where $\tilde{p}$ is a quasiparticle contribution to the pressure.

\begin{figure}[h]
\begin{center}
\includegraphics[scale=1]{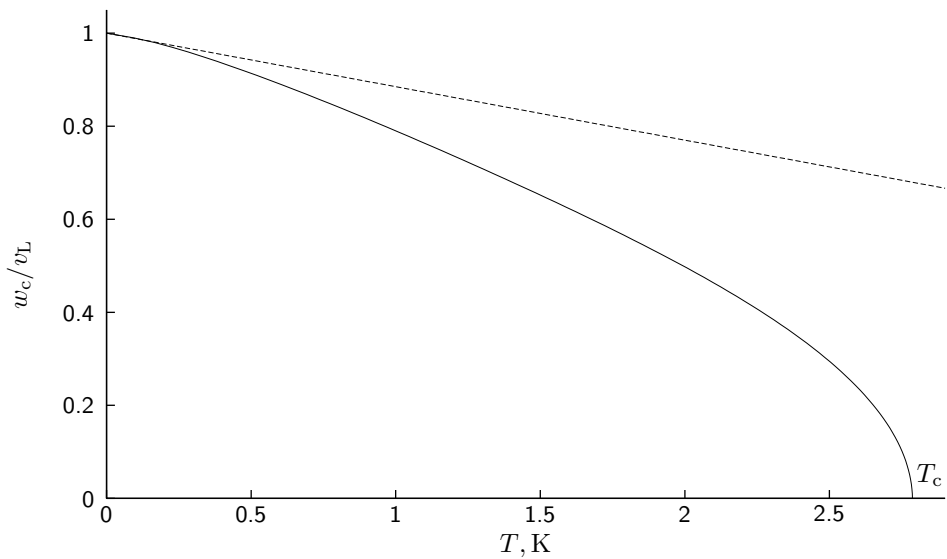}
\end{center}
\caption{Critical velocity $w_\text{c}$ versus temperature $T$ at normal
pressure. Dashed line corresponds to the equation $T=\Delta-p_0 w$. Note
that the condition $T<\Delta-p_0 w$ holds true over entire stability
domain. The ``stability'' critical velocity $w_\text{c}$ coincides with
the Landau critical velocity $v_\text{L}$ at zero temperature and
vanishes completely \eqref{lambdapoint} at the critical temperature
$T_\text{c}$ (the $\lambda$-point). In the
phonon-roton model the critical temperature is  $T_\text{c}\approx
2.8\,\text{K}$. } \label{crit}
\end{figure}

Inequality \eqref{M4b} is the first to become invalid. Appropriate validity region is plotted
in Fig.~\ref{crit}. The liquid is unstable above the curve.

At zero temperature the critical velocity becomes the Landau critical velocity $v_\text{L}$.
It should also be noted that
for systems
where all quasiparticles can be described hydrodynamically
(in other words, systems lacking roton branch) inequality \eqref{M4b} at zero temperature
includes $\pdcl{p}{\rho}{T,w}-w^2>0$ \ie $w<c$.

\section{Conclusion}
Experimentally, the superfluidity break-up in small orifices is believed to have the following nature
(see \cite{exp}).
Until the aperture size is too small the critical velocity does not depend on the temperature
and increases as the size decreases. This is the very behaviour that is specific to the vortex-related critical velocity.

When the orifice width is narrow enough the vortex-related critical velocity becomes so high,
that the break-up scenario and its features change. The critical velocity does not depend on the aperture any
more but decreases when the temperature increases. This behaviour is commonly associated
(see \cite{exp})
with the Iordanski-Langer-Fisher mechanism (see \cite{JLF}).
Nevertheless, this association lacks numerical comparison
because no reliable information about the actual orifice shape is available.

On the other side experimentally observed behaviour of the critical velocity can be attributed to the suggested
stability criterion. In other words we provide an alternative
explanation of experimental results based on an assumption that in narrow orifices
the thermodynamic limit of $w_\text{c}$ is reached.

We should also note that our approach to the critical velocity
as a stability limit is similar to that used by Kramer\cite{kramer}. Actually the
inequality he employed is not a thermodynamic one. Moreover, generally speaking
it is wrong. But numerical results for the critical velocity
he obtained using the phonon-roton model do not deviate much from those plotted
in Fig.\ref{crit}.

\section{Acknowledgments}
It is a pleasure for us to thank I.A.Fomin for useful discussions.
This work was supported in parts by INTAS grant 01-686,
CRDF grant RP1-2411-MO-02,
RFBR grant 03-02-16401, and RF president program.


\begin{thebibliography}{9}
\bibitem{exp}
{{E.Varoquaux, W.Zimmermann, and O.Avnel}, in
\textit{Excitations in Two-Dimensional and Three-Dimensional
Quantum Fluids}, edited by A.F.G.Wyatt and H.J.Lauter (NATO ASI Series,
Plenum Press, New York-London, 1991), p.343.}
\bibitem{khalat}
{I.M.Khalatnikov, \textit{An Introduction to the
theory of superfluidity}. (W.A.Benjamin, New York-Amsterdam
1965).}
\bibitem{LL5}
{L.D.Landau, E.M.Lifshitz, \textit{Statistical Physics},
part 1 (Pergamon Press, Oxford, 1980).}
\bibitem{don76}
{R.J.Donnelly, P.H.Roberts, \textit{Journal of Low Temperature Physics},
\textbf{27}, 687 (1977).}
\bibitem{don81}
{R.J.Donnelly, J.A.Donnelly, R.N.Hills, \textit{Journal of Low Temperature Physics},
\textbf{44}, 471 (1981).}
\bibitem{JLF}
{J.S.Langer, J.D.Reppy, \textit{Prog. Low. Temp. Phys.}, Vol. VI, ed. C.J.Gorter
(North-Holland, Amsterdam, 1970) Ch.1.}
\bibitem{kramer}
{L.Kramer, \textit{Phys. Rev.}, \textbf{179}, 149 (1969).}
\end{thebibliography}
\end {document}